\documentclass[a4paper]{jpconf}
\usepackage{graphicx}
\begin{document}
\title{Descope of the ALIA mission}

\author{Xuefei Gong$^{1}$, Yun-Kau Lau$^{1,2,3,4}$, Shengnian Xu$^{1}$, {Pau Amaro-
\newline Seoane$^{5}$}, Shan Bai$^{6}$, Xing Bian$^{1,7}$,
 Zhoujian Cao$^{1,3,4}$, Gerui Chen$^{8}$, Xian Chen$^{5}$,
Yanwei Ding$^{9}$, Peng Dong$^{1}$, Wei Gao$^{1,7}$, Gerhard Heinzel$^{10}$,
  Ming Li$^{9}$, Shuo Li$^{11}$, Fukun Liu$^{12,13}$, Ziren
Luo$^{14}$,  Mingxue Shao$^{2}$, Rainer Spurzem$^{11,13,15}$, Baosan Sun$^{16}$, Wenlin Tang$^{1,7}$, Yan
Wang$^{10}$, Peng Xu$^{2,17}$,  Pin Yu$^{18}$, Yefei Yuan$^{19}$,
 Xiaomin Zhang$^{8}$ and Zebing Zhou$^{16}$}

\address{$^1$ Institute of Applied Mathematics, Academy of Mathematics and Systems Science, Chinese
Academy of Sciences, No.55 Zhongguancun Donglu, Beijing 100190, China}

\address{$^2$ Morningside Center of Mathematics, Chinese Academy of Sciences,  No.55 Zhongguancun Donglu, Beijing 100190, China}

\address{$^3$ State Key Laboratory of Scientific and Engineering Computing, Academy of Mathematics and Systems Science, Chinese Academy of Sciences, No.55 Zhongguancun Donglu, Beijing 100190, China}

\address{$^4$ State Key Laboratory of Theoretical Physics£¬Institute of Theoretical Physics, Chinese Academy of Sciences, No.55 Zhongguancun Donglu, Beijing 100190, China}

\address{$^5$ Max-Planck-Institut f\"{u}r Gravitationsphysik (Albert Einstein Institut), Am Mf\"{u}hlenberg 1, D-14476 Potsdam, Germany}

\address{$^6$ Theoretisch-Physikalisches Institut,
Friedrich-Schiller-Universit$\hbox{\"{a}}$t Jena,
Fr$\hbox{\"{o}}$belstieg 1, D-07743 Jena, Germany}

\address{$^7$ University of Chinese Academy of Sciences,  No.19A Yuquan Road, Beijing 100049, China}

\address{$^8$ College of Applied Sciences, Beijing University of Technology, 100 Pingleyuan,
Beijing 100124, China}

\address{$^9$ Dongfanghong Small Satellite Company, F12 Shenzhou Building, No.31 Zhongguancun Nandajie, Beijing 100081, China}

\address{$^{10}$ Max-Planck-Institut f\"{u}r Gravitationsphysik (Albert Einstein Institut), Callinstra{\ss}e 38, D-30167 Hannover, Germany}

\address{$^{11}$ National Astronomical Observatories, Chinese Academy of Sciences,
20A Datun Road, Beijing 100012, China}

\address{$^{12}$ Department of Astronomy, School of Physics, Peking University, No.5 Yiheyuan Road, Beijing 100871, China}

\address{$^{13}$ The Kavli Institute for Astronomy and Astrophysics, Peking University, 5 Yiheyuan Road, Beijing 100871, China}

\address{$^{14}$ Institute of Mechanics, Chinese Academy of
Sciences, 15 Beisihuanxi Road, Beijing 100190, China}

\address{$^{15}$ Astronomisches
Rechen-Institut, Zentrum f\"{u}r Astronomie, Universit\"{a}t
Heidelberg, M\"{o}nchhofstrasse 12-14, 69120 Heidelberg, Germany}

\address{$^{16}$ School of Physics, Huazhong University of Science and Technology, 1037 Luoyu Road, Wuhan 430074, China; Center for Gravitational Experiments, Huazhong University of Science and Technology, 1037 Luoyu Road, Wuhan 430074, China}

\address{$^{17}$ Institute of Theoretical Physics, Chinese Academy of Sciences, No.55 Zhongguancun Donglu, Beijing 100190, China}

\address{$^{18}$ Mathematical Science Center, Tsinghua University, Beijing 100084, China}

\address{$^{19}$ Center for Astrophysics, Department of Astronomy,
University of Science and Technology of China, 96 Jinzhai Road,
Hefei 230026, China}

\ead{lau@amss.ac.cn}

\begin{abstract}
The present work reports on a feasibility study commissioned by the
Chinese Academy of Sciences of China to explore various possible
mission options to detect gravitational waves in space alternative
to that of the eLISA/LISA mission concept. Based on the relative
merits assigned to science and technological viability, a few
representative mission options descoped from the ALIA mission are
considered. A semi-analytic Monte Carlo simulation is carried out to
understand the cosmic black hole merger histories and the possible
scientific merits of the mission options  in probing the light seed
black holes and their coevolution with galaxies in early Universe.
The study indicates that, by choosing the armlength of the
interferometer to be three million kilometers and shifting the
sensitivity floor  to around one-hundredth Hz, together with a very
moderate improvement on the position noise budget, there are certain
mission options capable of exploring light seed, intermediate mass
black hole binaries at high redshift that are not readily accessible
to eLISA/LISA, and yet the technological requirements seem to within
reach in the next few decades for China.
\end{abstract}

\section{Introduction}
The present work originated from a recently completed second phase of the feasibility study
commissioned by National Space Science Center, Chinese Academy of Sciences of China to explore
various possible mission options to detect gravitational waves in
space alternative to that of (e)LISA mission concept. At the beginning
of the study dating back to more than three years ago, the ALIA (Advanced
Laser Interferometer Antenna) mission concept first proposed by Peter Bender \cite{BenderALIA,BenderBegelman} was
chosen as the starting point. It is conceivably in many ways the
simplest adaptation of the LISA mission concept to a measurement band
centered around a few hundredth Hz. A more detailed
study \cite{8thLISA} of the possible sciences of the mission  further
indicates that, apart from the known LISA sources, the mission also holds
the promise in mapping out the mass and spin distribution of
intermediate mass black holes (IMBHs) possibly present in dense star
clusters at low redshift as well as in shedding important light on
the structure formation in the early Universe. However, when the key
technologies of the mission is further looked at, the sub-picometer
interferometry requirement in the laser metrology part poses a major
obstacle on the technological side of the mission. With a view that
China  will have
a reasonable chance to realise the mission in the next few decades
and to minimise possible risks in future R\&D of the key
technologies, the task of mission descope is inevitably forced upon
us.

Upon further evaluation  of the relative merits between science and technological
viability of a  few representative descope options descended from the ALIA mission, it emerges that there exists certain class of mission design
that is technologically viable within the next few decades for China, and yet it contains significant science that goes beyond eLISA/LISA.
The principal aim
of this report is to outline this set of  mission design parameters and
briefly sketch the scientific case study.

The outline of the present work may be described as follows.
 In Section 2, we
will state the mission design parameters, display the corresponding sensitivity curves and the detection ranges with respect to
 black hole binaries with various mass ratios. In Section 3, a semi-analytic Monte-Carlo simulation is carried out to understand the scientific potential
 of the prospective missions
in probing the structure formation in early Universe.

\section{Mission descope}

After some careful considerations, the following set of baseline design
parameters will be chosen for future study and development.
\pagebreak
\begin{table}[h]
\begin{center}
\caption{\label{parameters}}
\begin{tabular}{lllll}
\br Armlength   &\ Telescope      &\ Laser
&1-way position &\ Acceleration\\

\ \ \ \ ($\textrm{m}$) & diameter ($\textrm{m}$)  &power($\textrm{W}$)
&\ noise
($\frac{\textrm{pm}}{\sqrt{\textrm{Hz}}}$)
&\ \ \ \ \ \ ($\frac{{\textrm m}~{\textrm s}^{-2}}{\sqrt{\textrm{Hz}}}$) \\

\mr $3\times10^9$ &\ \ 0.45-0.6   & \ \ \ 2
&\ \ \ \ \ \ \ 5-8
&$3\times10^{-15}$$(>0.1\textrm{mHz})$\\

\mr $5\times10^8$ (ALIA) &\ \ \ \ \ 1.0  & \ \ 30 &\ \ $\ \ \ \ \
0.1$
&$3\times10^{-16}$$(>1\textrm{mHz})$\\

\mr $5\times10^9$ (LISA) &\ \ \ \ \ 0.4   & \ \ \ 2
&\ \ \ \ \ \ \ 18
&$3\times10^{-15}$$(>0.1\textrm{mHz})$\\

\mr $1\times10^9$ (eLISA) &\ \ \ \ \ 0.2   & \ \ \ 2
&\ \ \ \ \ \ \ 11
&$3\times10^{-15}$$(>0.1\textrm{mHz})$\\

\br
\end{tabular}
\end{center}
\end{table}
\par
\begin{figure}[h]
\begin{center}
%\begin{minipage}{10pc}
\includegraphics[scale=0.5]{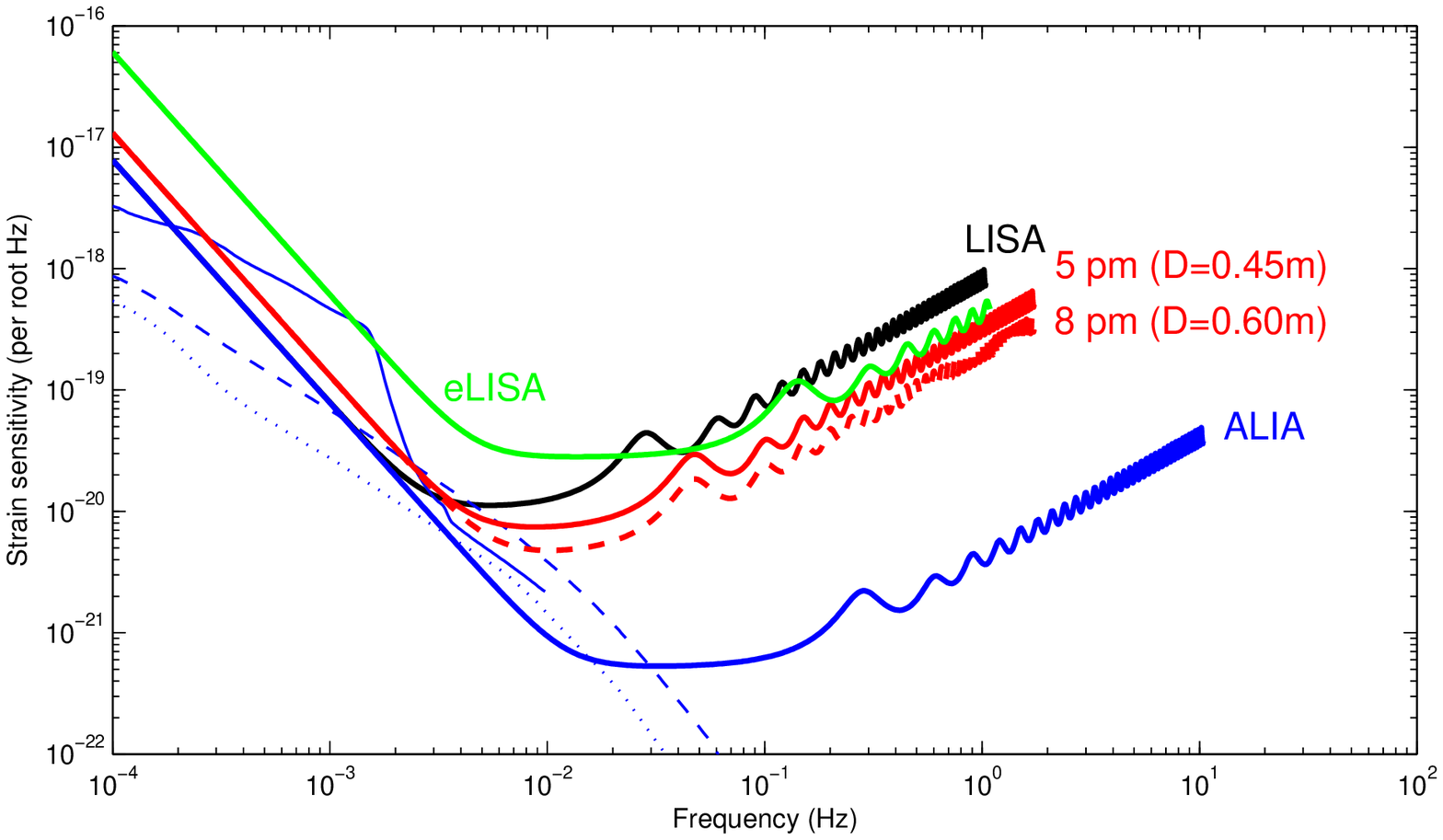}
\caption{ }\label{sensitivity}
%\end{minipage}\hspace{2pc}%
\end{center}
\end{figure}

\noindent
For reference purpose, the baseline design parameters of ALIA,
LISA/eLISA are also given in Table \ref{parameters}. The relevant sensitivity
curves are displayed in Figure \ref{sensitivity}. Apart from the
instrumental noises, confusion
 noise generated by both galactic and extra-galactic compact binaries are also taken into consideration. Relevant confusion levels are converted from estimations by Hils and Bender \cite{HilsBender} and Farmer and Phinney \cite{FarmerPhinney}.

\begin{figure}[h]
\begin{center}
%\begin{minipage}{10pc}
\includegraphics[width=26pc]{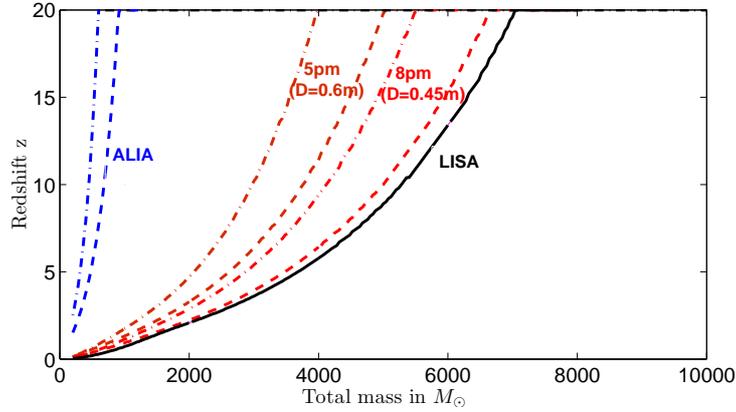}
\caption{All-angle averaged detection range under a
single Michelson threshold SNR of 7 for 1:4 mass ratio IMBH-IMBH
binaries, one year observation prior to merger. For each mission option, both upper and lower confusion noise levels (represented by the dashed curve and dotted dashed curve respectively) due to extragalactic compact binaries are considered.}\label{range1}
%\end{minipage}\hspace{2pc}%
\end{center}
\end{figure}

For black hole binaries with mass ratio $1:4$, typical of
what one would expect from hierarchical black hole growth at high redshift,
 the all angle averaged detection range are plotted in Figure \ref{range1}.
Apart from galactic confusion noise \cite{HilsBender}, upper level (dashed curve) and lower level (dotted dashed curve) of
 confusion noise generated by extragalactic compact binaries as those estimated by \cite{FarmerPhinney} are also taken into  account.

In calculating the averaged SNR, we have used  hybrid waveforms in
the frequency domain with black hole spin not taken into account \cite{Ajith_hybrid,Ajith_hybrid_2}.
For one year of observation before merger, the contributions in SNR
due to large spin is indeed negligible according to our
calculations. Spin is relevant only in the parameter estimation
stage, which will not be discussed in the present work. As may be
seen from Figure \ref{range1}, for a given redshift, the proposed mission concept
is capable of detecting lighter black hole binaries in comparsion
with eLISA/LISA and thereby provides better understanding of the
hierarchical assembling process in early Universe.

Apart from IMBH binaries at high redshift, the designed sensitivity
at around 0.01Hz measurement band means that the instrument is also
capable of detecting IMRIs (Intermediate Mass Ratio Inspirals)
harboured at globular clusters or dense young star clusters at low
redshift ($z<0.6$). See \cite{Pau_IMRI} for a further discussion of
the capture dynamics of an IMRI in dense star clusters. Displayed in
Fig 3 are the detection ranges of IMRIs with different mass ratios
one year prior to merger. The stellar black hole inspiral into an
IMBH is fixed to be $10M_{\odot}$, while the mass of the IMBH is
subject to variation to generate different mass ratios in the
figure.

\begin{figure}[h]
\begin{center}
%\begin{minipage}{10pc}
\includegraphics[width=26pc]{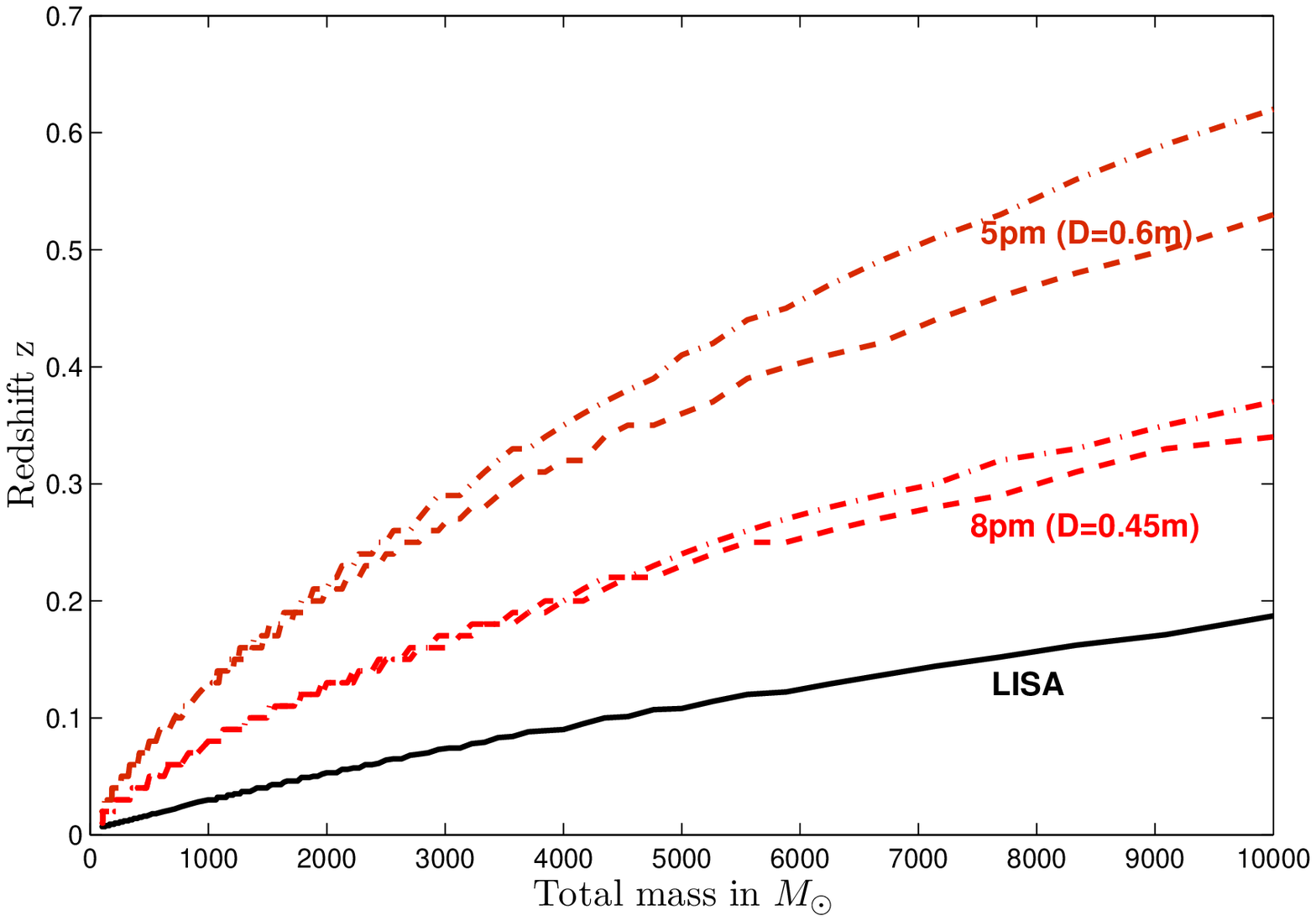}
\caption{All-angle averaged detection range under a
single Michelson threshold SNR of 7 for a stellar mass black hole
spiralling into IMBHs with reduced masses of $10M_{\odot}$, one year observation prior to merger. For each mission option, both upper and lower confusion noise levels (represented by the dashed curve and dotted dashed curve respectively) due to extragalactic compact binaries are considered.}\label{range2}
%\end{minipage}\hspace{2pc}%
\end{center}
\end{figure}

\section{Scientific case study}
The  primary science driver of the ALIA mission  is to make direct
detection of IMBH binaries at high redshift
descended from the heavy Pop III stars \cite{8thLISA,MadauRees,HegerWoosley} and thereby provide
insight into the physics of the structural formation processes at
early Universe.
 It would be of interest to find
out to what extent this science objective is compromised in the
descoped mission. For this purpose, we carry out a simple Monte
Carlo simulation of black hole merger histories based on the
realizations of EPS formalism and semi-analytical dynamics, in
accordance with the prescription given in
\cite{Volonteri2003_1,Volonteri2003_2,Sesana2007}.

\begin{figure}
\begin{center}
\includegraphics[width=3.6in,angle=0]{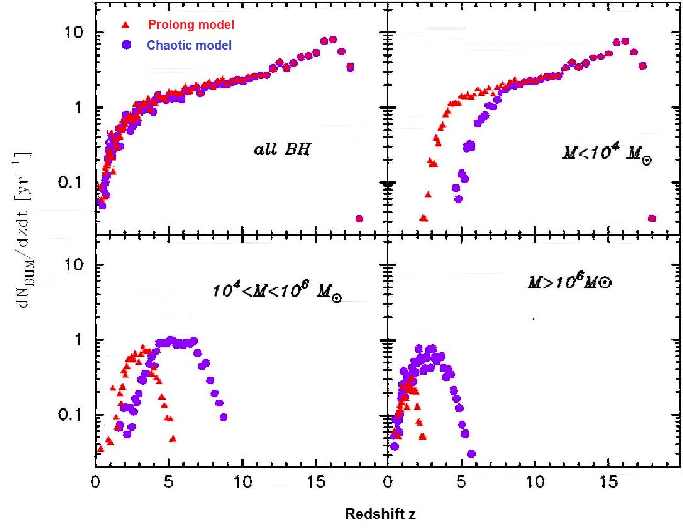}
\end{center}
\caption{Coalescence rate predicted by the realized simulations. }
\label{coalescencerate}
\end{figure}

\begin{figure}
\begin{center}
\includegraphics[scale=0.55]{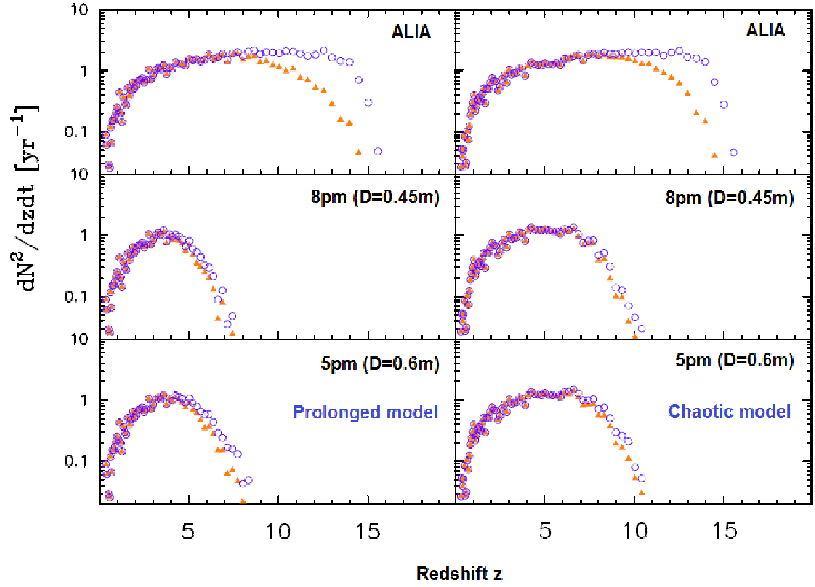}
\end{center}
\caption{Event rates for descoped options.}
\label{detectionrate}
\end{figure}

Pop III remnant black holes of $150M_{\odot}$ are placed in
$3.5\sigma$ biased halos at z=20 with initial spins of the seeds
generated randomly. By prescribing VHM-type
dynamics \cite{Volonteri2003_1,Volonteri2003_2}, we trace downwards
the black hole merging history. The halo mass ratio criteria for
major merger is set to be $>0.1$. Both the prolonged accretion and
the chaotic accretion scenario are considered. Black hole spins
coherently evolve through both mergers and accretions processes and
their magnitudes influence strongly the mass-to-energy conversion
efficiency. We assume efficient gaseous alignment of the black holes
so that the hardening time is short and only moderate gravitational
radiation recoils take place. Numerical simulations \cite{Khan_1}
suggest that
 the hardening and merging times scales remain short even in gas
 free environment.
 In
calculations relevant to GW observations, we assume a threshold SNR
of 7 for detection in the sense of single Michelson interferometer
and
 one year observation prior to merger.
The results are schematically given in Figure \ref{coalescencerate} and Figure \ref{detectionrate}.

We assess our simulations by fitting the black hole mass functions
and luminosity functions at six almost equally divided successive
redshift intervals ranging from z=0.4 to 2.1. In the
prolonged accretion scenario,
 the  results deviate from the observational
constrains given by Soltan type argument when going up to redshift
$z>1.5$. It may therefore
underestimate the black holes growth rate and perhaps also the
coalescence rate. Observationally the existence of very high
redshift ($z>6$) AGNs implies that feed back mechanisms may be very
different at early epoch so that fast growth of the seed black hole
could be possible.

In terms of coalescence rate, our result  displayed in Figure \ref{coalescencerate} is in
overall agreement with the results given by Sesana et al \cite{Sesana2007,Bayesian_1} and Arun
et al \cite{Arun}, though the coalescence counts given by their simulations
are about two or three times higher. It is likely due to  various
numerical discrepancies  in the simulations. Overall, our black hole
mass growth is slower, particularly in the prolonged accretion
scenario. At $z=15$, the total mass of the black hole binaries typically
are still less than $600M_{\odot}$ in the prolonged model and this
may lead to a smaller counts in detectable sources.
We expect our results  yields a very conservative (pessimistic)
estimate of black hole binaries merger event rate.

The astrophysics encapsulated in our simulation represents the state
of art understanding of structural formation after the dark age. Due
to our poor understanding of the evolution of the Universe at this
epoch, it is likely that the simulation overlooks many details of
the physical processes involved. The event rate count should be
looked upon in a cautious way. Instead of reading into the precise
numbers, it serves as an indication what spaceborne gravitational
wave detector is capable of and in our case, the advantage of setting
the most sensitive regime of the measurement band from a few mHz to
0.01Hz.

\subsection*{Globular cluster harbored IMRIs}

Before we conclude, let us also briefly estimate the detection
capability for IMRIs  in dense star clusters. The calculations and
underlying hypotheses are identical to that  in \cite{8thLISA}(see
also \cite{Will,Pau_birth}).
 The results are given in Table \ref{IMRIrate}.

\begin{table}[h]
\begin{center}
\caption{\label{IMRIrate}Prospective detection rate for IMRIs in
globular clusters.}
\begin{tabular}{ll}
\br Mission option   &\ \ Upper level of confusion  \ \   Lower level of confusion   \\

% & \ \ \ \ \ \ \ \ \ \ \ \ \ \ \ ($\times\frac{f_{tot}}{0.1}\frac{\nu_0}{10^{-10}{\textrm{yr}}^{-1}}\frac{10M_{\odot}}{\mu}$ $\textrm{yr}^{-1}$)  \\

\mr ALIA \ \ \ \    &$z_{c}=5$: \ \ \ $\sim8000$ \ \ \ \ \ \ \ \ \ \ \ \ \ \ \ \ \ \ \ \ \ $\sim12000$\\
 & $z_{c}=3$: \ \ \ $\sim6000$ \ \ \ \ \ \ \ \ \ \ \ \ \ \ \ \ \ \ \ \ \ $\sim7000$\\

%\mr 1L &\ \ \ \ \ \ \ \ \ \ \ \ \ $\sim500$ \ \ \ \ \ \ \ \ \ \ \ \ \ \ \ \ \ \ \ \ \ \ \ $\sim860$   \\

%\mr 1H &\ \ \ \ \ \ \ \ \ \ \ \ \ $\sim110$ \ \ \ \ \ \ \ \ \ \ \ \ \ \ \ \ \ \ \ \ \ \ \ $\sim140$   \\

\mr 5pm (D=0.6m) &\ \ \ \ \ \ \ \ \ \ \ \ \ $\sim90$ \ \ \ \ \ \ \ \ \ \ \ \ \ \ \ \ \ \ \ \ \ \ \ \ \ $\sim130$   \\

\mr 8pm (D=0.45m)&\ \ \ \ \ \ \ \ \ \ \ \ \ $\sim26$ \ \ \ \ \ \ \ \ \ \ \ \ \ \ \ \ \ \ \ \ \ \ \ \ \ $\sim32$   \\

\mr LISA &\ \ \ \ \ \ \ \ \ \ \ \ \ \ \ \ \ \ \ \ \ \ \ \ \ \ \ \ $\sim3$\ \ \ \ \ \ \ \ \ \ \ \ \ \ \ \ \ \ \    \\
\br
\end{tabular}
\end{center}
\end{table}

The above event rate estimate is subject to many uncertainties and perhaps we should not attach too much
importance to the precise numbers. Instead, the calculations serves as an indication
of the detection potential of the mission concept as far as IMRIs at low redshift are concerned.
Further, as event rate goes up as the
cubic of the improvement in sensitivity, it also brings out the
advantage of shifting slightly the most sensitive region of the
measurement band to a few hundredth Hz, as far as detection of IMRIs
is concerned. It should also be remarked that collision of dense
star clusters \cite{Pau_collision} constitutes a possible IMBH
gravitational wave sources, while the inspiral of massive black
holes ($\sim 10^3M_\odot$ to $\sim 10^4M_\odot$) into the
supermassive black hole at the center of a galaxy is also a
promising IMRI source \cite{Miller, fukun}. However, the
corresponding event rates would be difficult to estimate.

\pagebreak

\section{Concluding remarks}

With the second phase of the feasibility study of gravitational wave detection in space  coming to a close, a preliminary mission design deemed suitable as blue print for future development of the project in the Chinese Academy of Sciences is put forward.
 Together with the roadmap to  advance the project step by step, the mission design will serve as a guide for future developments on both the theoretical as well as technological fronts. We hope to report upon further progress of the project on various areas soon.

\section*{Acknowledgments}
The present work is part of the report submitted to the National Space Science Center,
Chinese Academy of Sciences for the project  entitled ``Feasibility study of gravitational wave detection in space''
(project number XDA04070400). Prof. Wenrui Hu's relentless effort to promote gravitational wave detection in space
in China motivated the feasibility study in the first place.
 We are also grateful to Prof. Shuangnan Zhang for his helpful suggestions and advices on many occasions
 throughout the course of our work. As always, Prof. Peter Bender generously shares with us his expertise
 and insights in this subject. Professors Shing-Tung Yau and Lo Yang have been very supportive and the
 Morningside Center of Mathematics provides a very conductive research environment to carry out the study.
 Partial support from the NSFC (contract numbers 11305255 and 11171329) and the Marine Public Welfare Project
 of China (contract number 201105032) are acknowledged. RS is supported by Chinese Academy of Sciences Grant
 Number $2009S1-5$ (RS), and through the "Thousand Talents" (Qianren) program of China.

\end{document}